\documentstyle[leqno,twoside,11pt]{article}

\catcode`\@=11

\def\footnotesize{\@setsize\footnotesize{11pt}\ixpt\@ixpt
       \abovedisplayskip \z@
      \belowdisplayskip\z@
     \abovedisplayshortskip\abovedisplayskip
    \belowdisplayshortskip\belowdisplayshortskip
\def\@listi{\leftmargin\leftmargini \topsep 3pt plus 1pt minus 1pt
     \parsep 2pt plus 1pt minus 1pt
    \itemsep \parsep}}

\skip\footins 12pt plus 12pt

\def\footnoterule{\kern3\p@  \hrule width 3em\vspace{3pt}} 

\def\ps@plain{\let\@mkboth\@gobbletwo
     \def\@oddfoot{{\hfil\small\thepage\hfil}}%
     \def\@oddhead{}
      \def\@evenhead{}\def\@evenfoot{}}

\def\ps@headings{\let\@mkboth\markboth
        \def\@oddfoot{}\def\@evenfoot{}%
        \def\@evenhead{{\rm\thepage}\hspace*{2pc}{\sc\leftmark}\hfil}%
        \def\@oddhead{\hfil{\noindent\sc\rightmark}\hspace*{2pc}{\rm\thepage}}%

\def\ps@myheadings{\let\@mkboth\@gobbletwo
 \def\@oddfoot{}\def\@evenfoot{}%
 \def\@oddhead{\hfil{\sc\rightmark}\hspace*{2pc}{\normalsize\rm\thepage}}%
 \def\@evenhead{{\normalsize\rm\thepage}\hspace*{2pc}{\sc\leftmark}\hfil}%
}}

\def\abstract{\if@twocolumn
\section*{Abstract}
\else \small
\begin{center}
{\bf Abstract\vspace{-.5em}\vspace{3pt}}
\end{center}
\quotation
\fi}
\def\endabstract{\if@twocolumn\else\endquotation\fi}

\def\newproof#1{\@nprf{#1}}

\def\@nprf#1#2{\@xnprf{#1}{#2}}

\def\@xnprf#1#2{\expandafter\@ifdefinable\csname #1\endcsname
\global\@namedef{#1}{\@prf{#1}{#2}}\global\@namedef{end#1}{\@endproof}}

\def\@prf#1#2{\@xprf{#1}{#2}}

\def\@xprf#1#2{\@beginproof{#2}{\csname the#1\endcsname}\ignorespaces}

\def\newalgorithm#1{\@ifnextchar[{\@oalg{#1}}{\@nalg{#1}}}

\def\@nalg#1#2{%
\@ifnextchar[{\@xnalg{#1}{#2}}{\@ynalg{#1}{#2}}}

\def\@xnalg#1#2[#3]{\expandafter\@ifdefinable\csname #1\endcsname
{\@definecounter{#1}\@addtoreset{#1}{#3}%
\expandafter\xdef\csname the#1\endcsname{\expandafter\noexpand
  \csname the#3\endcsname \@thmcountersep \@thmcounter{#1}}%
\global\@namedef{#1}{\@alg{#1}{#2}}\global\@namedef{end#1}{\@endalgorithm}}}

\def\@ynalg#1#2{\expandafter\@ifdefinable\csname #1\endcsname
{\@definecounter{#1}%
\expandafter\xdef\csname the#1\endcsname{\@thmcounter{#1}}%
\global\@namedef{#1}{\@alg{#1}{#2}}\global\@namedef{end#1}{\@endalgorithm}}}

\def\@oalg#1[#2]#3{\expandafter\@ifdefinable\csname #1\endcsname
  {\global\@namedef{the#1}{\@nameuse{the#2}}%
\global\@namedef{#1}{\@alg{#2}{#3}}%
\global\@namedef{end#1}{\@endalgorithm}}}

\def\@alg#1#2{\refstepcounter
    {#1}\@ifnextchar[{\@yalg{#1}{#2}}{\@xalg{#1}{#2}}}

\def\@xalg#1#2{\@beginalgorithm{#2}{\csname the#1\endcsname}\ignorespaces}
\def\@yalg#1#2[#3]{\@opargbeginalgorithm{#2}{\csname
       the#1\endcsname}{#3}\ignorespaces}

\def\@beginproof#1{\rm {\it #1.\ }}
\def\@endproof{\outerparskip 0pt\endtrivlist}

\def\@begintheorem#1#2{\it {\sc #1\ #2.\ }}
\def\@opargbegintheorem#1#2#3{\it
      {\sc #1\ #2\ (#3).\ }}
\def\@endtheorem{\outerparskip 0pt\endtrivlist}

\def\@beginalgorithm#1#2{\rm \trivlist \item[\hskip \labelsep{\sc #1\ #2.}]}
\def\@opargbeginalgorithm#1#2#3{\rm \trivlist
      \item[\hskip \labelsep{\sc #1\ #2.\ (#3)}]}
\def\@endalgorithm{\outerparskip 6pt\endtrivlist}

\newskip\outerparskip

\def\trivlist{\parsep\outerparskip
  \@trivlist \labelwidth\z@ \leftmargin\z@
  \itemindent\parindent \def\makelabel##1{##1}}

\def\@trivlist{\topsep=0pt\@topsepadd\topsep
  \if@noskipsec \leavevmode \fi
  \ifvmode \advance\@topsepadd\partopsep \else \unskip\par\fi
  \if@inlabel \@noparitemtrue \@noparlisttrue
    \else \@noparlistfalse \@topsep\@topsepadd \fi
    \advance\@topsep \parskip
  \leftskip\z@\rightskip\@rightskip \parfillskip\@flushglue
  \@setpar{\if@newlist\else{\@@par}\fi}%
  \global\@newlisttrue \@outerparskip\parskip}

\def\endtrivlist{\if@newlist\@noitemerr\fi
   \if@inlabel\indent\fi
   \ifhmode\unskip \par\fi
   \if@noparlist \else
      \ifdim\lastskip >\z@ \@tempskipa\lastskip \vskip -\lastskip
         \advance\@tempskipa\parskip \advance\@tempskipa -\@outerparskip
         \vskip\@tempskipa
   \fi\@endparenv\fi
   \vskip\outerparskip}

 \newproof{@proof}{Proof}

 \newtheorem{@theorem}{Theorem}[section]
 \newenvironment{theorem}{\begin{@theorem}}{\end{@theorem}}

\newproof{Example}{Example}
\newproof{Method}{Method}
\newproof{Exercise}{Exercise}

 \def\@figtxt{figure}
\long\def\@makecaption#1#2{\small
\setlength{\parindent}{18pt}
\baselineskip 14pt
 \ifx\@captype\@figtxt
 \vskip 10pt
 \setbox\@tempboxa\hbox{{\sc #1} {\it #2}}
 \ifdim \wd\@tempboxa >\hsize {\sc #1} {\it #2}\par \else \hbox
to\hsize{\hfil\box\@tempboxa\hfil}%
 \fi\else\hbox to\hsize{\hfil{\sc #1}\hfil}%
 \setbox\@tempboxa\hbox{{\it #2}}%
 \ifdim \wd\@tempboxa >\hsize {\it #2}\par \else
 \hbox to \hsize{\hfil\box\@tempboxa\hfil}\fi
 \vskip 10pt
 \fi}

\def\fnum@figure{\par\sc Fig. \thefigure.\ }
\def\fnum@table{\small \sc Table \thetable}

\def\section{\@startsection {section}{1}{\z@}{-3.5ex plus -1ex minus
 -.2ex}
{2pt}{\large\bf}}
\def\subsection{\@startsection{subsection}{2}{\z@}{-3.25ex plus -1ex minus
 -.2ex}
{2pt}{\large\bf}}
\def\subsubsection{\@startsection {subsubsection}{3}{\z@}{1.3ex plus .5ex minus
    .2ex}{-.5em plus -.1em}{\normalsize\bf}}

\def\thebibliography#1{%
\parindent 0em
\vspace{9pt}
\begin{flushleft}\large\bf {References}\end{flushleft}
\addvspace{3pt}\nopagebreak\list
{[\arabic{enumi}]} {\settowidth\labelwidth{[#1]}
\leftmargin\labelwidth
\leftmargin=17pt
 \advance\leftmargin\labelsep
 \usecounter{enumi}\@bibsetup}
\def\newblock{\hskip .11em plus .33em minus -.07em}
 \sloppy\clubpenalty4000\widowpenalty4000
 \sfcode`\.=1000\relax}

\def\@bibsetup{
\itemsep=0pt \parsep=0pt
\small}

\def\theindex{\@restonecoltrue\if@twocolumn\@restonecolfalse\fi
\columnseprule \z@
\columnsep 35pt\twocolumn[\chapter*{Index}]
 \parskip\z@ plus .3pt\relax\let\item\@idxitem}

\def\printindex{\cleardoublepage\markboth{INDEX}{INDEX}
\addcontentsline{toc}{chapter}{Index}\@input{\jobname.ind}}

\ps@headings

\def\A{{\cal A}}
\def\G{{\cal G}}
\def\C{{\cal C}}
\def\AGb\overline{\A/\G}
\def\S{\Sigma}
\def\g{\gamma}

\def\div{{\rm div}}
\def\Gra{{\rm Gra}}
\begin{document}
\cleardoublepage
\pagestyle{myheadings}

\title{Differential Geometry for the Space of Connections Modulo
Gauge
Transformations\footnote{The  results presented in this paper
come from  the joint works of the author with Abhay Ashtekar
\cite{AL2}\cite {AL3}.}}
\author{Jerzy Lewandowski\thanks{Physics Department,  University
of Florida,
Gainesville, 32611 USA and IFT, Wydzial Fizyki, Uniwersytet
Warszawski,
ul Hoza 69, 00-681 Warszawa, Poland. Supported   by the National
Science
Foundation grant PHY91--07007 and by the Polish Committee for
Scientific
Research (KBN) through grant no.2 0430 9101}}

\date{April 1994}
\maketitle
\markboth{Jerzy Lewandowski} {Differential Geometry on
$\overline{\A/\G}$}

\pagenumbering{arabic}

This short article is a continuation of a longer, review
work, in the same volume by Ashtekar, Marolf and Mour\~ao \cite{AMM}.
The reader may find there an introduction to, and the summary of
the recent theory of diffeomorphism invariant measure for the
quotient space of connections $\A/\G$.  The measure is defined
on the projective limit $\overline {\A/\G}={\cal Q}$ (see (30)
\cite{AMM}) constructed from the loops in $\S$.
Associated with the measure on $\overline {\A/\G}$ is
the Hilbert space $H(\overline {\A/\G})$ of the squer integrable
functions. In the following
paper, we outline a new framework which may be called a
differential geometry for the projective limit $\overline {\A/\G}$.
 All
the details and other results are to be found
in joint papers of the author with Abhay Ashtekar \cite{AL2} and
\cite{AL3}.  The motivation for introducing the geometry is that
it provides us with powerfull techniques to construct various
operators defined in $H(\overline {\A/\G})$. In particular, we
introduce the commutator algebra of `vector fields' on $\overline
{\A/\G}$, define a divergence of a vector field and construct a
quantum
representation for them. Among the vector fields, there are
operators which we identify in [3,4] as regularised Rovelli-Smolin
\cite{RS}
loop operators  linear in momenta. A related class of operators are
the Ashtekar strip operators. They turn out to be expressible
in terms of the vector fields and get extended to
self adjoint operators [3,4]. We are also able to define (regularised)
higher order Rovelli-Smolin loop operators. Another class of
operators which comes out naturally are Laplace operators. We give
below a
 general
recipe and show two particular examples of a Laplace operator
associated to:
(i)a surface $s$ in  $\S$, (ii) a  metric tensor on $\S$.

Let us turn now to some details.  The key idea of our
construction  is glueing topologys, measures, differentiable manifolds,
 Hilbert spaces, operators etc. associated with the components of a
 certain projective
family $(\A_\g, \G_\g, \pi_{\g\g'})_{\g, \g'\in \Gra(\S)}$ introduced
below (see
also Baez \cite{B2}). The labeling set is the set $\Gra(\S)$ of all
the graphs in $\S$
(analytically embedded).  A given graph $\g$, $\A_\g$ is defined to
be the
space of connections restricted to  $\g$ and devided by the gauge
transformations which are constant on the vertexes of $\g$. $\A_\g$
is given
a differentiable manifold structure diffeomorphic to $G^E$, where
$E$ is the
number of the edges of $\g$.
The remainig gauge transformations form a compact Lie group  $\G_\g$
which acts
in $\A_\g$ and is isomorphic with $G^V$, $V$ denoting  the number of the
vertexes in $\g$. Given two graphs $\g,\g'\in\Gra(\S)$ we say that
$\g'\geq\g$,
whenever the immage of $\g'$ containes that of $\g$ and the
vertexes of $\g$ coincide with the vertexes of $\g'$. Whenever
$\g'\geq\g$ then, there is defined a  natural map (projection)
$\pi_{\g\g'}: (\A_{\g'}, \G_{\g'}) \rightarrow (\A_\g, \G_\g)$. The map
$\pi_{\g\g'}$ is a smooth surjection of $\A_{\g'}$ onto $\A_{\g}$ and a
smooth homomorphism of $\G_{\g'}$ onto $\G_{\g}$. Moreover,
\begin{equation}
\pi_{\g\g'}\circ\pi_{\g'\g''}=\pi_{\g\g''}.
\end{equation}
This completes the definition of the projective family $(\A_\g, \G_\g,
\pi_{\g\g'})_{\g, \g'\in \Gra(\S)}$. The family privides us with the
projective
limits $\overline{\A}$ and $\overline{\G}$ respectively (we follow here
Marolf\&Mour\~ao \cite{MM}). $\overline{\G}$ is a compact
Hausdorff group and acts on $\overline{\A}$. The quotient satisfies
\cite{AL2}
\begin{equation}
\overline{\A}/\overline{\G} = {\cal Q}
\end{equation}
where ${\cal Q}$ is the projective limit of the quotients
$(\A_\g/\G_\g)_{\g\in\Gra(\S)}$ and coincides with the projective
limit (30) in
\cite{AMM}.

We are at the position to begin the glueing procedure. The space of {\it
$C^n$  functions on $\overline{\A}$}, $\C^n(\overline{\A})$, is defined
to be
\begin{equation}\label{gl}
\C^n(\overline{\A}):= \bigcup_{\g\in\Gra(\S)}C^n(\A_\g)\ /\sim
\end{equation}
where $f_\g\in C^n(\A_\g)$ is $\sim$ -- equivalent to  $h_{\g'}\in
C^n(\A_{\g'})$ whenever for any $\g''\geq\g,\g'$
\begin{equation}
\pi_{\g\g''}^*f_\g=\pi_{\g\g'}^*h_{\g'}.
\end{equation}
$\C^0(\overline{\A})$ is in a 1-to-1 correspondence with the space of the
cylindrical
functions on $\A$ (extended to $\overline{\A}$). (Substituting
$C^n(\A_{\g})$
in (\ref{gl})
with the space of differential $k$--forms on $\A_\g$ we define {\it
differential
$k$--forms } $\Omega^k(\overline{\A})$.)

{\it A vector field $X$ on} $\overline{\A}$ is defined to be a family
$(X_\g)_{\g\geq\g_0}$,
where each $X_\g$ is a vector field on $\A_\g$, $\g_0$ is a fixed
graph and
\begin{equation}\pi_{\g\g'*}X_{\g'}=X_\g,
\end{equation}
whenever $\g'\geq\g\geq\g_0$. We equipe each $\A_\g$ with a natural
probability measure $\mu_\g$ induced via $\A_\g=G^E$ from the
product measure
given by a fixed Haar measure on $G$. The family of measures
$(\mu_\g)$ is
self consistent and defines the natural measure on $\overline{\A}$
(corresponding to
the measure $\mu^0$ (see \cite{AMM}).
A vector field $X=(X_\g)_{\g\geq\g_0}$ on $\overline{\A}$ is
{\it compatible with the measure $\mu^0$} if for every $\g,\g'\geq \g_0$
\begin{equation}
\div X_\g\sim\div X_{\g'}.
\end{equation}
Then, the divergences  $\div X_\g$ define a smooth  function on
$\overline{\A}$
denoted by $\div X$.

The vector fields on $\overline{\A}$  compatible with the natural
 measure
$\mu^0$
form a commutator algebra and
\begin{equation} \label{div}
\div [X,Y]=X(\div Y) - Y(\div X).
\end{equation}

The  group $\overline{\G}$ acts naturally on vector fields on
$\overline{\A}$.
 An important statement is that {\it every
 $\overline{\G}$--invariant vector
  field on $\overline{\A}$ is compatible
with the natural measure $\mu^0$}.

The Hilbert space $H(\overline{\A/\G})$ of cylindrical functions
 is the subspace
 of
$\overline{\G}$--invariant elements of $L^2(\overline{\A},\mu^0)$.
 A $\overline{\G}$--invariant vector field $X$ on $\overline{\A}$
  defines an
operator acting on a dence domain $\C^1(\overline{\A/\G})\subset
H(\overline{\A/\G})$ consisting of the $\overline{\G}$--invariant
 elements of $\C^1(\overline{\A})$.  It
follows from the properties of the divergence of $X$ that the
hermitian
conjugate to $X$ is, as in a finite dimensional case, given by
the following
formula
\begin{equation}
X^\star = -X -\div X.
\end{equation}
A {\it quantum momentum operator} corresponding to $X$ may be
defined to be
\begin{equation}
P(X):={i\over 2}(X-X^\star)= i(X+{1\over 2}\div X)
\end{equation}
Here is our main result:

\begin{theorem}
Let $X$ be a $\overline{\G}$--invariant vector field on
$\overline{\A}$ and
$(P(X), \C^1(\overline{\A/\G}))$ be
the corresponding  momentum operator; then $P(X)$ is symmetric
and extendable
 to a self adjoint operator; let $Y$ be also a
 $\overline{\G}$--invariant
 vector
field on $\overline{\A}$; then,
\begin{equation}
[P(X),P(Y)]=iP([X,Y]).
\end{equation}
\end{theorem}

We  turn now to the Laplace operator. Since we have thus far so
succesfully eploited
the Lie group geometry of $G$, it is natural to fix an invariant metric
tensor
on $G$ and attempt to apply the product geometry of $\A_\g=G^E$.
However, this Riemann structure
of $\A_\g$ is not consistent with the projections $\pi_{\g\g'}$ and the
resulting
family of the Laplace operators wouldn't be well defined. A solution to
this problem  which
we offer, is to assigne  to each finite analytic curve $e$ in $\S$
(a potential
edge of a graph) a number  $l(e)$ (a `length')
which is additive; i.e.,
\begin{equation}
l(e_1\circ e_2)= l(e_1)+l(e_2).
\end{equation}
Then, for each  graph $\g$ we define
\begin{equation}
\Delta^{(l)}_\g:= l(e_1)\Delta_{e_1}+...+ l(e_E)\Delta_{e_E}
\end{equation}
where $\Delta_{e_i}$ is a an operator which applied to a function
$f_\g(g_{e_1},...,g_{e_1})$ acts only on  the $G$-variable $g_{e_i}$ as
the Laplace operator for $G$.

\begin{theorem} The family of operators $(\Delta^{(l)}_\g,
C^2(\A_\g))_{\g\in
\Gra(\S)}$ is  consitent with the equivalence relation $\sim$ in the
following
sence: for every $f_\g\in C^2(\A_\g)$ and $f_{\g'}\in C^2(\A_{\g'})$
\begin{equation}
f_\g\sim f_{\g'}\ \  \Rightarrow\ \  \Delta^{(l)}_{\g'}f_{\g'} \sim
\Delta^{(l)}_{\g}f_\g;
\end{equation}
the operator $\Delta^{(l)}$ acting on a domain $\C^2(\overline{\A/\G})$ by
\begin{equation}
\Delta^{(l)}([f_\g]_\sim):=[\Delta^{(l)}_\g(f_\g)]_\sim
\end{equation}
is symmetric and extendable to a self--adjoint operator.
\end{theorem}

\noindent
{\bf Examples of the laplace operators.}

(i) Suppose, there is given a  surface $s$ in $\Sigma$. Define
$l(e)$ to be the number of isolated points of intersection between
$e$ and $s$.

(ii) Let $g$ be a metric tensor in $\S$. Define $l(e)$ to be the
length of $e$.

In particular, when we consider in the Example (i) $s$ being a point
in $\S$,
the obtained Laplace operator takes on the appearance of a regularised
operator ${\delta \over \delta A^i_a(s)}{\delta \over \delta A^i_a(s)}$.

\end{document}